\documentclass[namedreferences]{solarphysics}

\usepackage[hyperref,optionalrh,showbiblabels]{spr-sola-addons} 
\usepackage{graphicx}        
\usepackage{color}           
\usepackage{breakurl}        

\chardef\us=`\_

\begin{document}

\begin{article}
\begin{opening}

\title{Sources of The Slow Solar wind During the Solar Cycle 23/24 Minimum} 

\author{E.K.J.~\surname{Kilpua}$^{1}$\sep
M. S.~\surname{Madjarska}$^{2}$\sep
N.~\surname{Karna}$^{3}$\sep
T.~\surname{Wiegelmann}$^{4}$\sep
C.~\surname{Farrugia}$^{5}$\sep
W.~\surname{Yu}$^{5}$\sep
K.~\surname{Andreeova}$^{6}$\sep
}

\runningauthor{Kilpua {\it et al.,}}
\runningtitle{Slow solar wind origin}

\institute{$^{1}$ Department of Physics, P.O. Box 64, University of Helsinki, Helsinki, Finland \\ email: \url{Emilia.Kilpua@helsinki.fi} \\
$^{2}$ Armagh Observatory, College Hill, Armagh BT61 9DG, UK \\
$^{3}$ NASA Goddard Space Flight Center, Greenbelt, MD, USA; George Mason University, Fairfax, VA, USA \\
$^{4}$ Max-Planck-Institut feur Sonnensystemforschung, Justus-von-Liebig-Weg 3, D-37077 Goettingen, Germany \\
$^{5}$ University of New Hampshire, 8 College Road Durham, NH 03824-3525, US \\
$^{6}$ Finnish Meteorological Institute, P.O. Box 503, Helsinki, Finland}

\begin{abstract}
We investigate the characteristics and the sources of the  slow ($< 450 km s^{-1}$) solar wind during the four years (2006--2009) of low solar activity between Solar Cycles 23 and 24. We use a comprehensive set of {\it in-situ} observations in the near-Earth solar wind (Wind and ACE) and remove the periods when large-scale interplanetary coronal mass ejections were present.  The investigated period features significant variations in the global coronal structure, including the frequent presence of low-latitude active regions in 2006--2007, long-lived low- and mid-latitude coronal holes in 2006-- mid-2008 and mostly the quiet Sun in 2009. We examine both Carrington Rotation averages of selected solar plasma, charge state and compositional parameters and distributions of these parameters related to Quiet Sun, Active Region Sun and the Coronal Hole Sun. While some of the investigated parameters (e.g., speed, the C$^{+6}/$C$^{+4}$ and He/H ratio) show clear variations over our study period and with solar wind source type, some (Fe/O) exhibit very little changes. Our results highlight the difficulty in distinguishing between the slow solar wind  sources based on the inspection of the solar wind conditions. 
\end{abstract}
\end{opening}


\section{Introduction}
The heliosphere is permeated by a continuous stream of charged particles emanating from the Sun's hot outer corona. This so-called ``solar wind'' carries the magnetic field of the Sun throughout the solar system and it is the medium where coronal mass ejections (CMEs), the key drivers of space weather storms in the near-Earth space environment, propagate. Moreover, the boundary conditions for space-weather modelling and forecasting are entirely dependent upon solar observations and, especially, on the knowledge of the origin of the solar wind. Whereas the regions of open magnetic field lines on the Sun, coronal holes, have been firmly established as the main source of the fast solar wind ($\sim700~km~s^{-1}$ \citealp{kri73}), the origin of the highly-structured slow ($\sim < 450~km~s^{-1}$) solar wind is still highly debated. 

The suggested sources for the slow solar wind are numerous, including: 1) fast converging open magnetic field lines near the boundaries of coronal holes \cite[e.g.,][]{wan90}; 2) transient plasma blobs that are released from the helmet streamers or pseudo-streamers \cite[e.g.,][]{wan98, wan00, she09, ril12}; 3) plasma released by reconnection between open and closed field lines at the coronal hole boundaries  \cite[e.g.,][]{mad04, fis05, lio05, mad09, edm12}; 4) hot outflows with speeds up to $\sim100~km~s^{-1}$ from the edges of  active regions, first reported by \citet[][]{koj99} and most recently by \citet[][and the references therein]{van2015}; and 5) jets originating from coronal bright points (BPs, small loop-like structures that are omni-present including coronal-holes and  the quiet-Sun) \cite[e.g.,][]{sub10, mad12,yuh2014,par10}. There is still no general consensus on whether all the above-mentioned processes can make a significant contribution to the slow solar wind and what is their relative importance in different phases of the solar activity cycle. However, considering the wide range of suggested sources, it is not surprising that the slow solar wind exhibits such variability of its properties \cite[e.g.][]{sch06}. 

The majority of the above-described sources assume that the plasma is intrinsically trapped in closed magnetic field lines where it is released via magnetic reconnection. Small plasma clouds and narrow jets are indeed detected on a daily basis in extreme ultraviolet (EUV) imaging and in spectroscopic \cite[e.g.,][]{dob00, sub10, mad12} and in white-light coronagraph observations \cite[e.g.,][]{wan98, woo99, dav09}. Furthermore, small-scale transients have been identified {\it in-situ} \cite[e.g.,][]{mol00, fen08, kil09, fou11, kil12, yu14}, most of them embedded in the slow solar wind. There is increasing evidence that the transient activity seen in different wavelengths in remote-sensing observations is linked. For instance, an extensive statistical analysis by \inlinecite{par10} confirmed the X-ray jets from BPs as the source of the narrow white-light ejections.  In addition, in a few cases coronal jets and plasma blobs have been traced to large distances from the Sun  using wide angle white-light heliospheric imaging \cite[e.g.,][]{rou11, yuh2014}, even up to their direct detection near the Earth \cite[][]{rou11}. However, the rate of small transients detected {\it in-situ} is significantly less (a few events per month) than the rate of jets and plasma blobs seen in remote-sensing observations (several tens per day). This suggests that the majority of small transient plasma blobs emitted from the Sun lose their identity before reaching the orbit of the Earth \cite[see discussion e.g. in][]{kil12}. 

 One way to establish the linkage is to compare elemental abundances and charge state properties at the Sun and in the solar wind \cite[e.g.][]{gle95, fel05}.  The charge states of the heavy ions (e.g., C$^{+6}/$C$^{+5}$, C$^{+6}/$C$^{+4}$, and O$^{+7}/$O$^{+6}$), freeze-in  in the corona when the solar wind expansion time-scale exceeds the ionization/recombination time-scales and thus provide an indicator of the coronal electron density and temperature profile in the freeze-in region \cite[e.g.,][]{zur99, zha09}. \cite{lan12} showed that the carbon charge states are the best indicators of the temperature of the corona and in particular the C$^{+6}/$C$^{+4}$ ratio is sensitive to the  solar wind type. However, using ion charge states to probe the solar wind origin is not straightforward. \cite{lep13} emphasized that charge states evolve considerably over a solar cycle, and hence, although they might be good proxies of the solar wind sources even across a given year, they cannot be used as an  absolute discriminator of the source.

In turn, the elemental ratios at the Sun depend in a  complex way on the chromospheric temperatures,  the magnetic field configuration where the plasma originates and the confinement time of the plasma in closed coronal loops. The spectroscopic remote-sensing observations have shown that newly emerging active regions loops tend to have similar abundances of elements with low   (e.g., Fe, Ne, Si, Mg)  and high (e.g., Fe, Ne, Si, Mg) first ionization potentials (FIP), while the abundance of high FIP elements increases with increasing confinement times \cite[e.g.][]{fel03,fel05}.

  ``FIP bias'' refers to the enhancement of elements with low first ionization potentials  (FIP) (e.g., Fe, Ne, Si, Mg) to those with higher FIP (e.g., O or S).  The FIP fractionation has recently been suggested to be driven by the ponderomotive force of the magnetohydrodynamic waves in the chromosphere and low corona, see e.g., \cite{lam15} and the references therein.  In the slow solar wind the abundances of elements with low FIP are typically larger by a factor of two to four compared to those with higher FIP \cite[e.g.,][]{zur06}. At the Sun strong FIP biases have been reported in active regions \cite[e.g.,][]{bak13}. In turn, for coronal streamers the results are more controversial and values vary from the core to the edge of the streamers \cite[e.g.][]{ray97, par00, bem03, uzz07}. Hence, it is not straightforward to make the association based on elemental abundances due to large variability in those values even in the same coronal source. Similar to heavy charge states, the solar wind composition exhibits a clear solar cycle trend, being considerably higher at solar maximum than during low solar activity periods  \citep{lep13}.


 
 
In addition, specific entropy $\ln{T_p/n_p^{\gamma-1}}$ has been shown to differentiate between the solar wind from different source regions \cite[e.g.,][]{pag04}. The fast solar wind has high temperatures, but low densities, and hence, generally high specific entropies. As mentioned above, for the slow solar wind  variations in plasma and magnetic field parameters are relatively large, and consequently, also the variations  in the specific entropy. The particularly low specific entropy structure in the slow solar wind is the high density and low temperature heliospheric plasma sheet (HPS).

In this paper we study the properties of the slow solar wind during the extended low solar activity period between Solar Cycles 23 and 24 (2006--2009). We examine a comprehensive set of solar wind parameters, including FIP fractionation, heavy ion charge states and specific entropy. The studied period of low solar activity is excellent for investigating the variations in the slow solar wind and its sources. The  contribution from large-scale CMEs was very small at this time \cite[e.g.,][]{kil12} and the global structure of the coronal magnetic field experienced drastic changes, featuring periods frequent with low-latitude coronal holes and active regions and an extended period of a very quiet Sun. The paper is organized as follows: In Section~2 we present our data and analysis methods. Section~3 gives the results of the statistical analysis. In Section~4 we discuss and summarize our results.

\section{Data and methods}
Our analysis combines 1-hour  averaged solar wind plasma and magnetic field observations from the Near-Earth Heliospheric data base (OMNI; \citealp{kin05}) and the charge state and elemental compositional characteristics (Fe/O and C$^{+6}/$C$^{+4}$) from the SWICS instrument of the ACE spacecraft. ACE was launched in August 1997 and it operates  close to the Lagrangian point L1. OMNI is a combination of L1 and near-Earth measurements (during our study period Wind and ACE data has been used) and the data has been shifted in time to the magnetopause. We obtained the OMNI data  through the NASA Goddard Space Flight Center Coordinated Data Analysis Web (CDAWeb, \url{http://cdaweb.gsfc.nasa.gov/}) and the ACE data from the ACE Science Center (\url{http://www.srl.caltech.edu/ACE/ASC/level2/}). The reason behind using solar wind plasma magnetic field data from OMNI instead from ACE is that after 2006 there are significant  and persistent data gaps in the ACE solar wind density measurements.  We select solar wind C$^{+6}/$C$^{+4}$ ratio to present the heavy ion charge states according to study by \cite{lan12} (see Introduction) and Fe/O to present the variations in the FIP bias. We also checked the analysis results using other charge states (C$^{+6}/$C$^{+5}$ and O$^{+7}/$O$^{+6}$), but no significant differences were obtained to those using the C$^{+6}/$C$^{+4}$ ratio.

To obtain only solar wind periods unperturbed by CMEs we have removed the interplanetary CME-intervals by combing the events from a published list in \cite{kil12} and from the online Richardson and Cane catalog (\url{http://www.srl.caltech.edu/ACE/ASC/DATA/level3/icmetable2.htm}). As discussed in Section~1 during our study period (2006--2009) only a handful of interplanetary CMEs were identified near L1. 
 
\begin{figure*}[t]
\vspace*{-5mm}
\begin{center}
\includegraphics[width=12.5cm]{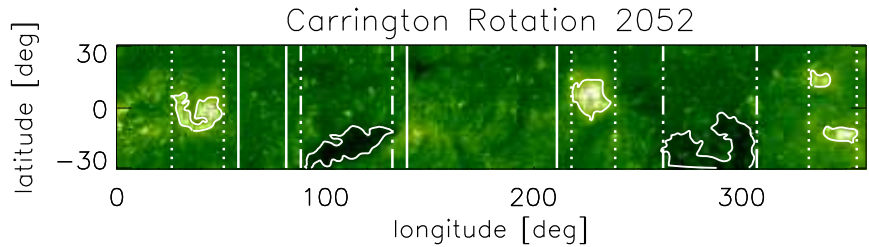}
\end{center}
\vspace*{-2cm}
\caption{A synoptic SOHO/EIT 195~\AA\ map for  Carrington Rotation 2052 (January 8, 2007 -- February 4, 2007). The dotted vertical lines bound the periods when active regions were present (Active Region Sun: ARS), the dash-dotted lines select the coronal holes periods (Coronal Hole Sun: CHS) and the dashed lines mark the quiet Sun (QS) periods. We have required that these periods are separated by half a day (7$^{\circ}$).}
\label{fig:synoptic}
\end{figure*}

We used synoptic maps produced from full-disk images taken in the 195~\AA\ channel of  the Extreme-ultraviolet Imaging Telescope \citep[EIT,][]{del1995} to calculate the fractional  active-region and coronal-hole areas for the CR~2039 starting on 2006 January 18 to CR~2091 ending on 2010 January 3. The preparation of the synoptic maps  from CR~2039 to CR~2055 is given in \cite{ben2001} (made by E. Benevolenskaya, hereafter EB) and from CR~2056 to CR~2092 in \cite{hess2014} (by N. Karna, NK). To determine the contour levels that determine the AR and CH areas, we first calculated the background emission as the mean of each individual Carrington map for pixels with Digital Numbers (DN) below 1700~DN/s (the pixels  above 1700 DN/s are associated with saturated signal rather than real signal most probably caused by cosmic rays not removed by the standard  data reduction procedure). The background emission is then obtained as the mean plus 1.1 times the standard deviation of the data for each Carrington map. Similar approach was successfully used by \citet{mad09} and \citep{sub10}. For the contours of the AR areas we used a value of 1.3 times the background emission for the Carrington maps produced  by  EB  and 1.5 times the  background emission for the maps  of  NK. The reason for the  different thresholds is the slightly different resolution of the two datasets of maps. All steps of the analysis were accompanied by a careful visual inspection of the goodness of the  contour levels.  The coronal-hole contour levels are  0.1$\times$BG for the EB maps and 0.23$\times$BG for the NK maps. Examples of the AR and CH contours are given in Figure~\ref{fig:synoptic}.

First, we calculated from the CR EIT synoptic maps the relative fractional ares of  the Active Region Sun (ARS) and  the Coronal Hole Sun (CHS) using the definitions described above. Since the photospheric footpoints of the field lines that map to the ecliptic can have a wide latitudinal range, we calculated these fractions using a  latitude range of $\pm 30^{\circ}$. We also checked the results using various other latitudinal ranges up to  $\pm 60^{\circ}$. This resulted in essentially similar solar cycle variations in the ARs and CHS fractions. We also investigated in more detail the properties of the slow ($< 450~km~s^{-1}$) solar wind associated with different sources, i.e. QS, ARS and CHS. For this part of the study we estimated the in-situ periods that are affected by the QS, ARS, CHS by defining the boundaries of these periods from the synoptic maps, see an example in Figure \ref{fig:synoptic}. The corresponding times in the near-Earth solar wind are found by assuming that the solar wind propagates radially at a constant speed. We use the average slow solar wind speed for our whole dataset, 359 $km~s^{-1}$ (corresponding the Sun to Earth transit time of 166.7 hours).  

\begin{figure}[ht]
\vspace*{2mm}
\begin{center}
\includegraphics[width=8cm]{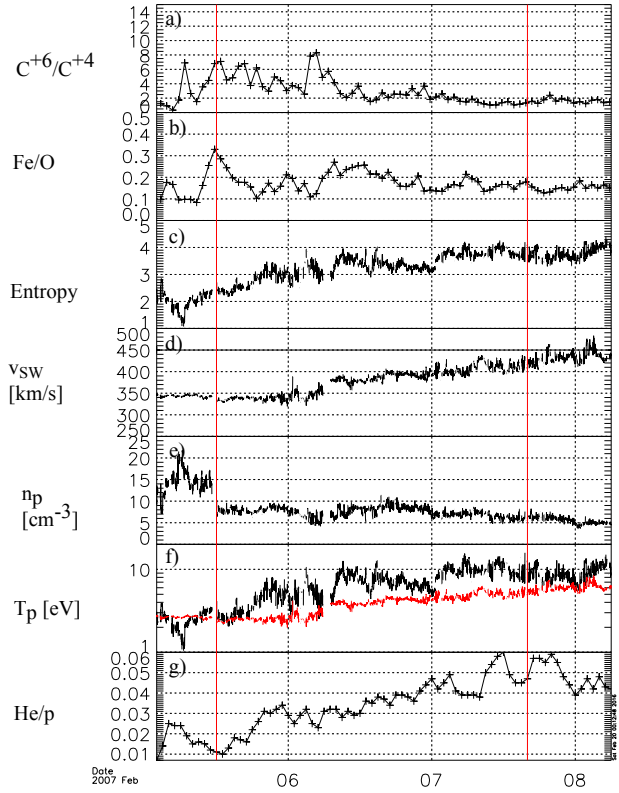}
\end{center}
\caption{The solar wind measurements during 2007 February 5, 13~UT -- 2007 February 7, 15~UT, corresponding the active region interval in Figure~\ref{fig:synoptic}. The panels show from top to bottom:  a) C$^{+6}/$C$^{+4}$ ratio, b) Fe/O ratio, c) specific entropy, d) speed, e) density, f) temperature and g) Helium to proton ratio.}
\label{fig:example_SW}
\end{figure}

For instance, the dotted lines  that bound the left most active region in Figure \ref{fig:synoptic} correspond to the period of 2007 January 31, 21~UT -- 2007 February 2, 19~UT. The corresponding in-situ interval is 2007 February 5, 13~UT -- 2007 February 7, 15~UT and is bounded in Figure \ref{fig:example_SW} by red lines. We require that  the ARS (bounded by the dotted lines), the CHS  (bounded by the dash-dotted lines) and the QS periods (bounded by the solid lines) are separated from each other by at least half a day (i.e., $\sim 7^{\circ}$ in Carrington maps). The total hours of different sources are distributed as follows: The QS periods cover 7070 hours (294.6 days), the AR periods -- 1886 hours (53.6 days), and the CH periods 395 hours (16.5 days). Figure \ref{fig:example_SW} also illustrates that solar wind properties may change considerably in the slow solar wind over relatively short time intervals (about three day interval shown).

\section{Statistical results}

\begin{figure}[t]
\vspace*{2mm}
\begin{center}
\includegraphics[width=8.3cm]{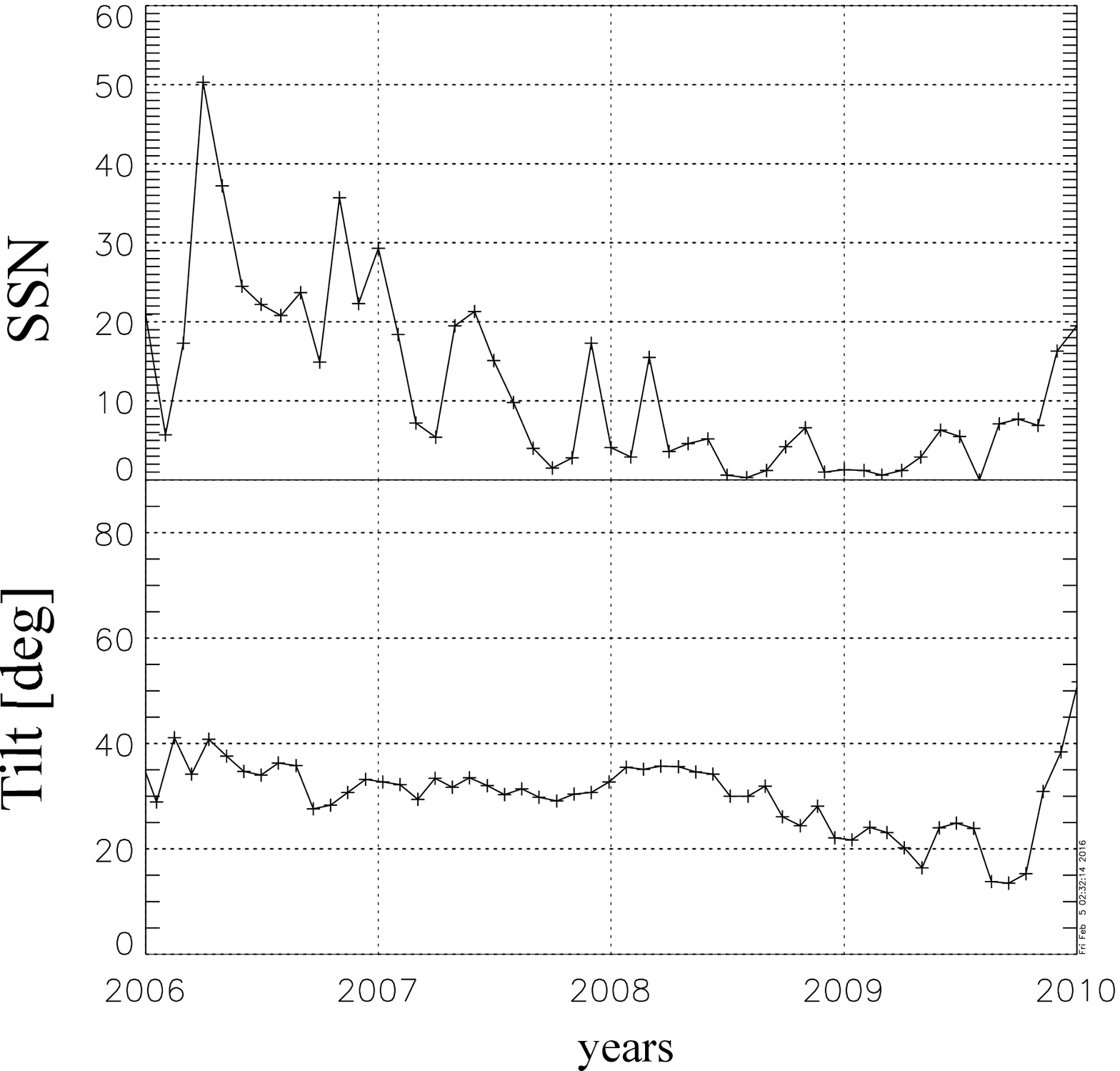}
\end{center}
\caption{Top) Monthly smoothed sunspot number from Solar Influences Data Center, bottom)  Maximum extent of the neutral line of the heliospheric current sheet for each Carrington Rotation from the Wilcox Solar Observatory.}
\label{fig:SN}
\end{figure}

The monthly sunspot number from the Solar Influences Data Center (\url{http://sidc.oma.be}) and the maximum latitudinal extend of the neutral line of the heliospheric current sheet (HCS) from the Wilcox Solar Observatory (\url{http://wso.stanford.edu}) are shown in Figure~\ref{fig:SN}. The top panel illustrates that solar activity was relatively low during our whole study period, in particular during 2008--2009 when for most of the time the visible disk of the Sun was completely ``spotless''. The bottom panel of Figure~\ref{fig:SN} shows that the HCS neutral line tilt reached low values, i.e. the Sun's magnetic field was dipole-like,  only towards the end of 2009. This implies that during most of the extended low activity period between Solar Cycles 23 and 24 the solar magnetic field had significant multipole components \cite[e.g.,][]{abr10}. See also Figure 5 in  \cite{kil11}, which shows the synoptic maps for the HCS neutral line for the selected Carringon Rotations for 2007--2010 demonstrating that  between 2007 and mid 2008 the HCS neutral line experienced significant warps and extended to high latitudes.

The flat and low latitude configuration  was reached only in mid 2009 and in late 2009 the HCS neutral line suddenly shifted to higher latitudes (seen also in Figure~\ref{fig:SN}) and become more warped. These may also affect the ecliptic slow solar wind. During the times when the HCS neutral line is flat and confined to low latitudes the ecliptic slow wind is largely influenced  by  the main streamer belt, while during the times  when the HCS neutral line is warped and has high tilt,  significant contributions from coronal holes and their boundary regions are expected.

\begin{figure*}[ht]
\vspace*{2mm}
\begin{center}
\includegraphics[width=12cm]{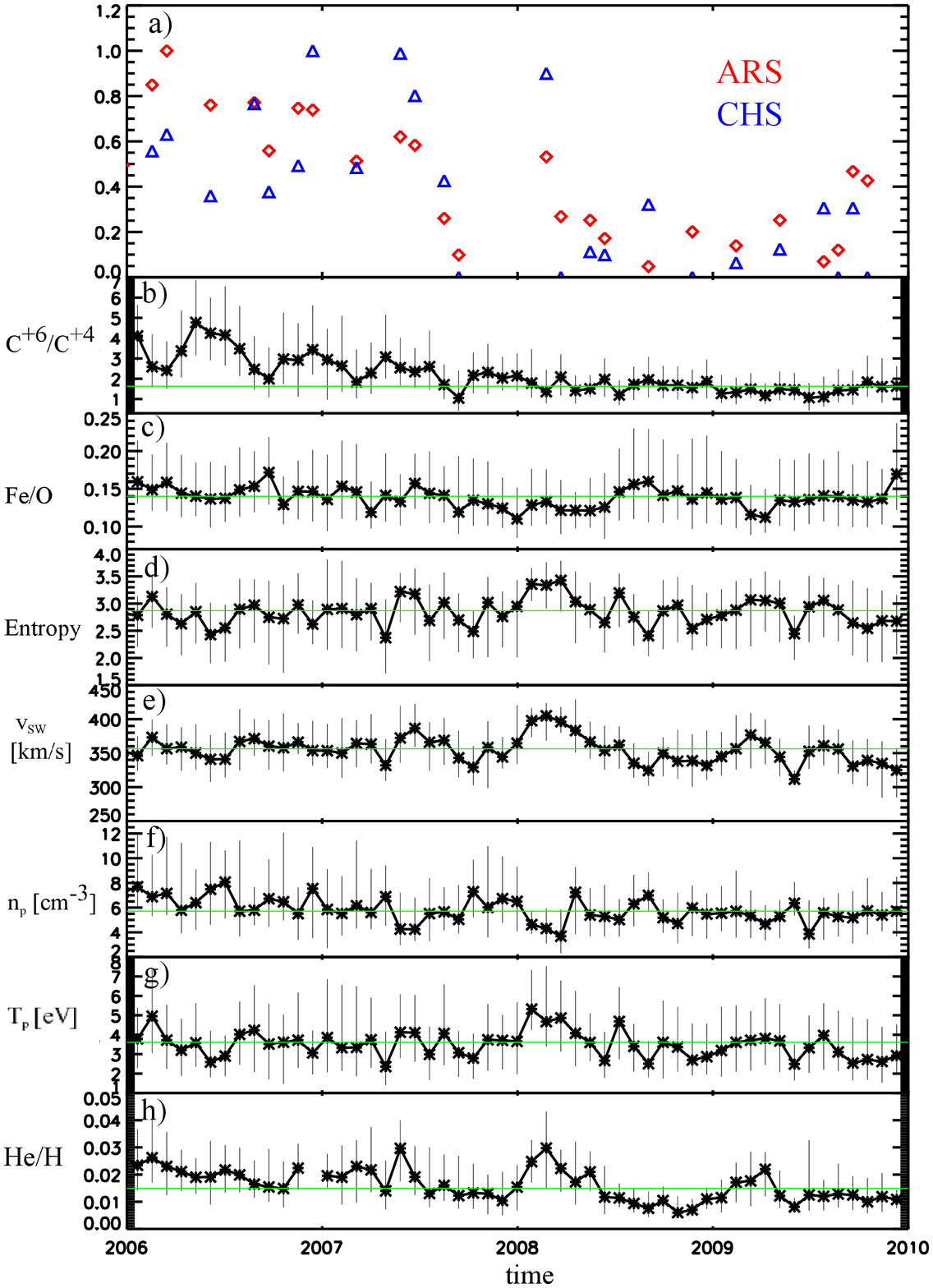}
\end{center}
\caption{The medians of selected solar wind parameters calculated over Carrington Rotations (27.2753 days) during our four year study period of 2006--2009 for the slow ($< 450$) solar wind. The error bars indicate the range from the lower to upper quartile. The green horizontal lines give the medians calculated using the whole investigated period. The panels show from top to bottom a) the normalized occurrence of the `Active Regions Sun'' (ARS; red diamonds), and ``Coronal Hole Sun'' (CHS; blue triangles), solar wind b) C$^{+6}/$C$^{+4}$ ratio, c) Fe/O ratio, d) specific entropy, e) speed, f) density, g) temperature, h) Helium to proton ratio.}
\label{fig:insituCR}
\end{figure*}

The top panel of Figure~\ref{fig:insituCR} shows the normalized (to the maximum value) CR fractions of the CHS and ARS, see Section 2 for the definitions. The variations in the number of active regions should follow the sunspot number and indeed the fraction of ARS was highest from the beginning of 2006 to early 2008, while the active regions were practically absent during the deepest solar minimum in 2008--2009. During the first half of our study period the identified active regions were primarily low-latitude active region related to the ``old'' cycle 23, while a few active regions detected in the end of 2009 mark the emergence of the mid-latitude active regions pertaining to the ``new'' cycle 24. The fraction of CHS was highest from the beginning of 2007 to mid 2008. The decrease in the fraction of low-latitude CHS is consistent with the Sun's magnetic field becoming increasingly dipolar by mid-2009 and having increasing effect from the main streamer belt (see discussion above). 

The medians of the selected solar wind parameters  calculated over each investigated Carrington Rotation (i.e. 27.2753 days) are displayed in panels b) - h) of Figure \ref{fig:insituCR}. As mentioned above, we have considered only those periods when the solar wind speed was $<450~km~s^{-1}$, and in addition, we have removed interplanetary CME intervals (see Section~2). The error bars show the ranges from the lower to the upper quartile, i.e. the interquartile range (IQR) describing the spread of the values. 

\begin{figure}[t]
\vspace*{2mm}
\begin{center}
\includegraphics[width=13.0cm]{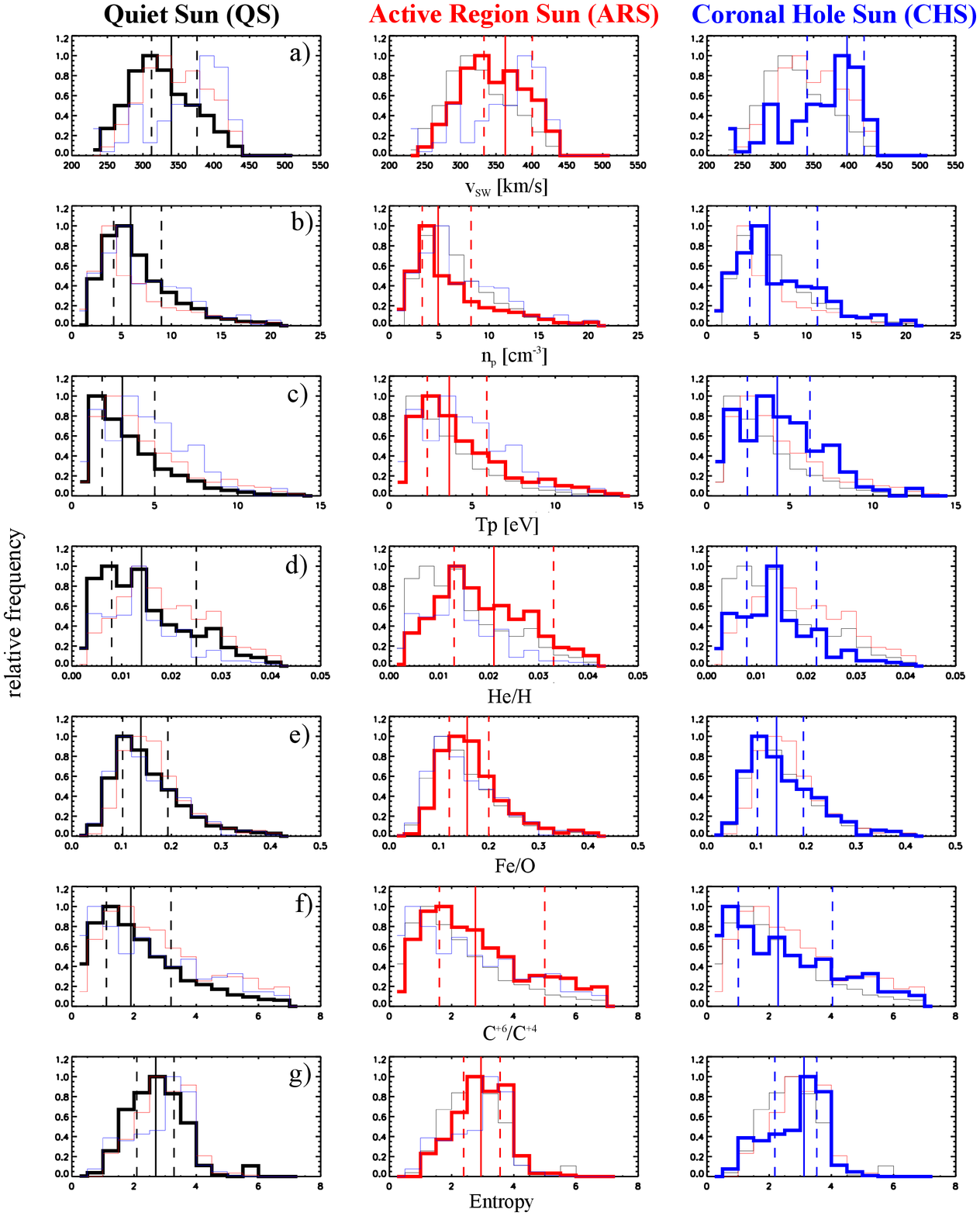}
\end{center}
\caption{Distributions of solar wind parameters for the periods of a) ``Quiet Sun'' (thick black line), b) ``Active Region Sun'' (thick red line), and c) ``Coronal Hole Sun'' (thick blue line). See Section 2 for the details. The thin curves in each panel show the other two distributions to help the comparison. The solid vertical lines show the medians and the dashed vertical lines mark the lower and upper quartiles. }
\label{fig:dist}
\end{figure}

The majority of the investigated solar wind parameters show quite modest and random variations over our four year study period, i.e. there is no clear correspondence with the variations of the CHS and ARS fractions. The solar wind from the mid-2008 to mid-2009 when both active regions and coronal holes were absent is the slowest wind and has the lowest He/H ratio. The year 2006 when the active regions were the most abundant is characterized by the largest C$^{+6}/$C$^{+4}$ ratio and the highest solar wind densities. The He/H ratio was also relatively high. The  period from mid-2007 to mid-2008 had relatively large contribution from low-latitude coronal holes. Note that the complete synoptic maps are mostly missing for the first half of 2008, but the inspection of the few existing EIT images reveals a significant presence of low latitude coronal holes during this period of time. The same signature  is also visible from Figure 2 of \cite{abr10} (which ends at  mid-2008). This time period is featured by the lowest Fe/O ratio, the fastest solar wind with the lowest densities and the highest temperature,  and the highest He/H ratio. 

Next, we  compare the solar wind properties between the periods when there were prominent low/mid-latitude coronal holes and active regions to periods of predominantly quiet Sun  (see Section 2 for the description of our approach). The results are shown in Figure \ref{fig:dist}. In the left panel of Figure \ref{fig:dist} the black thick lines show the distributions for the QS related slow solar wind periods. The distributions for the ARS and CHS related slow solar wind periods are shown with light red and light blue lines, respectively, to help the comparison between the different sources. Similar logic is used in the middle and right panels, but now the thick lines denote the ARS (red) and CHS (blue) related solar wind, respectively. The solid vertical lines indicate the medians and the dashed vertical lines the lower and upper quartiles of the given distributions. The medians and the lower and upper quartile ranges (i.e. the difference between the median and lower quartile and the upper quartile and median, respectively) are given in Table~1. 

Figure \ref{fig:dist}a and Table~1 show that the QS wind is clearly the slowest, while the fastest speeds are found for the CHS wind. Note also that the speed distribution for the CHS wind has relatively small upper quartile range, indicating that the CHS wind has a strong bias towards our slow solar wind  upper threshold  of 450~$km~s^{-1}$. The lower quartile for the CHS wind is $350~km~s^{-1}$ (i.e. only 25\% of the data had speeds below this value), which is very close to the  upper quartile for the QS wind, 370$~km~s^{-1}$.  The shape of the speed distribution and the median for the ARS wind are more similar to the QS wind than to the CHS wind. However, ARS wind speed distribution is flatter in contrast to QS and CHS wind distributions. 

The CHS and QS wind have larger mean densities than the ARS wind (Figure \ref{fig:dist}b). This is mainly due to the lack of the highest densities for the ARS related solar wind. The shapes of the density distributions are rather similar for all source types with strong clustering towards the lowest densities and an extended tail towards larger values, see also Table 1. The temperature distributions (panel  \ref{fig:dist}c) exhibit rather similar profiles as the density distributions. The CHS wind has also considerable higher temperatures than the QS and ARS related slow solar wind periods. The lowest temperatures are found for the QS solar wind. 

The next three panels (\ref{fig:dist}d--\ref{fig:dist}f) examine the solar wind charge state and compositional characteristics. The He/H distributions and their medians differ quite considerably between different solar wind types. The ARS wind has considerably larger He/H median than the QS and CHS wind and its distribution features a long and extended tail.  The QS wind shows a strong clustering of the He/H values towards the lowest values. For the QS and CHS wind the upper quartiles are 0.023 and 0.022, respectively, close to the median for the ARS wind, 0.020.  The Fe/O ratio distributions are remarkably similar for all three source types and the medians and the lower and upper quartiles are also nearly identical. In turn, the C$^{+6}/$C$^{+4}$ ratio shows clear differences. This ratio is highest for the ARS wind and lowest for the QS wind. Unlike the Fe/O distribution which shows Gaussian-like shape, the C$^{+6}/$C$^{+4}$ distributions tend to peak at low values and have an extended tail. Note that while the QS wind lacks high C$^{+6}/$C$^{+4}$ values, the CHS wind has nearly identical fraction of highest values as the ARS wind.  The last panel of Figure \ref{fig:dist} gives the specific entropy. This parameter  shows the largest median for the CHS wind and the lowest median for the QS wind. However, the differences in the medians are not very large, but the  distributions have quite different shapes. The distribution peaks at high entropy values for the CHS wind while for the QS wind the distribution shape is more Gaussian-like.

\begin{figure}[t]
\vspace*{2mm}
\begin{center}
\includegraphics[width=8.0cm]{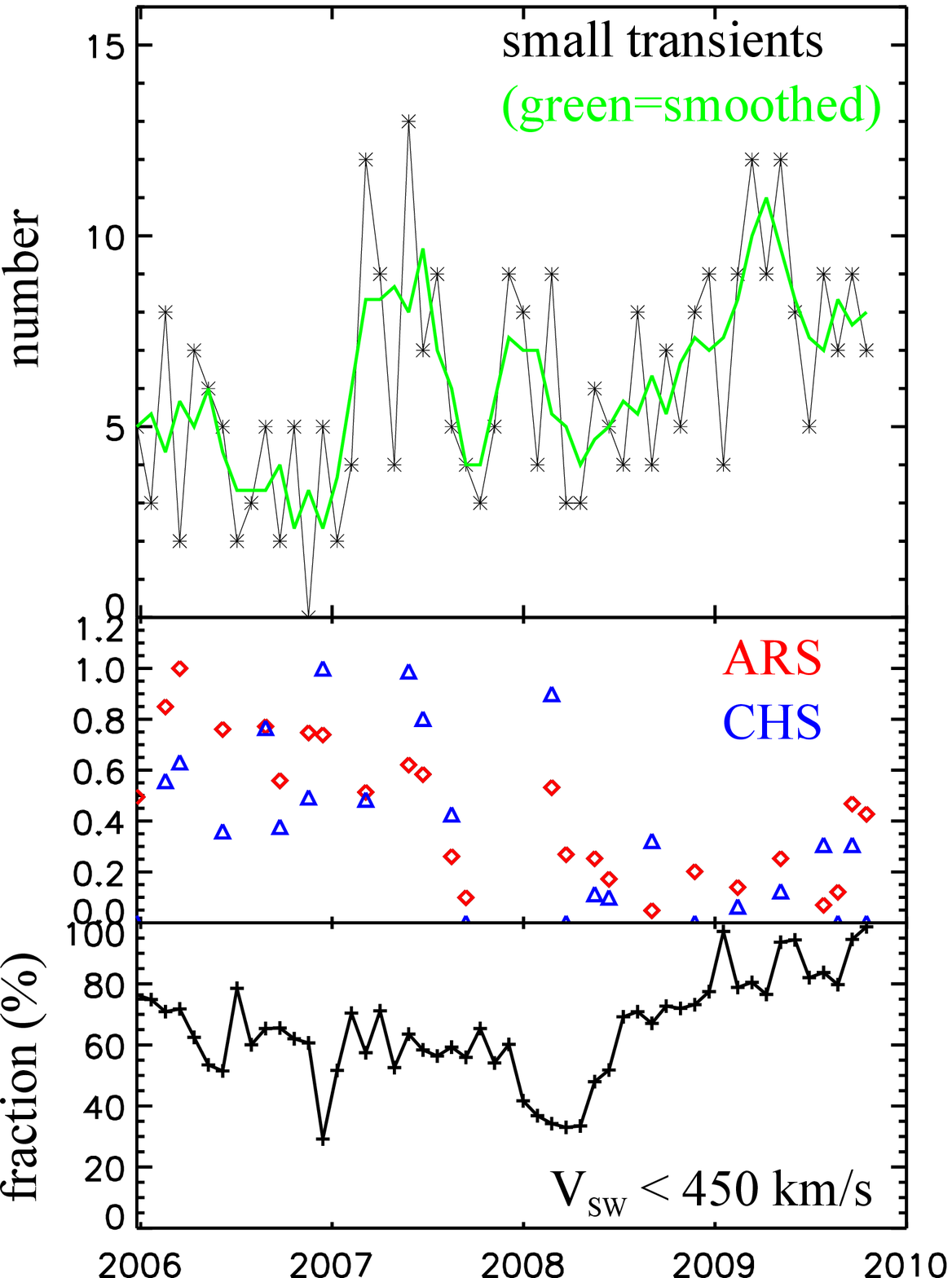}
\end{center}
\caption{Small transient contributions to the slow solar wind. The panels show: Top) the number of small transients identified in the slow solar wind for each Carrington rotation investigated (green line shows the smoothed number), Middle) the normalized occurrence of the `Active Regions Sun'' (ARS; red diamonds), and ``Coronal Hole Sun'' (CHS; blue triangles), i.e, same as in Figure 4, Bottom) the fraction of slow solar wind  for each investigated Carrington rotation.}
\label{fig:ST}
\end{figure}

The top panel of Figure \ref{fig:ST} shows the rate of ``small solar wind transients''  in the slow solar wind from the statistics published in \cite{yu14}. The green line is the smoothed number. Small transients are ``ICME-like'' coherent structures (often having a flux rope geometry) in the solar wind that have durations significantly less than large-scale CMEs ($< 12$ hours). As discussed in the Introduction, it has been suggested that the small transients would form a significant part of the slow solar wind. The rate of small transients was  the lowest in 2006, i.e. during the year with the highest active region contribution. The rate increased during the first half of 2007, i.e., when the  CHS contribution increased.  For the deepest solar minimum in 2008--2009 the number of small transients increased quite steadily until around September 2009 when the tilt of the neutral line of the HCS suddenly increased  (see Figure \ref{fig:SN}). Part of the variations in the number of small transients is presumably related to the variations in the fraction of slow solar (middle panel of Figure \ref{fig:ST}), which was only about 30-40\% during the first half of 2008 and reached nearly 100\% in 2009. However, the observed behaviour could also reflect the association of small transients with the ejections of small plasma blobs from streamers and/or from coronal holes boundaries. The last column of Table 1 gives the medians and the upper and lower quartial ranges for the same set of solar wind parameters that were calculated for the QS, ARS and CHS wind averaged over the small transients. The median C$^{+6}/$C$^{+4}$ values of small transients is similar to that of the QS wind, while speeds are more similar to ARS wind. 


\section{Discussion and Conclusions}
In this paper we have investigated the statistical properties of the slow ($< 450~km~s^{-1}$) solar wind associated with different solar sources. We studied both the Carrington Rotation medians and the distribution of various solar wind parameters connected to QS, ARS and CHS periods during the four years of low solar activity between 2006 and 2009. 

While some of the investigated solar wind parameters showed clear differences depending on the dominant source,  others had very little effect. For instance, the Carrington rotation medians for the solar wind FIP bias (here investigated using the Fe/O ratio) showed very little change over our four year study period. In addition, the distributions and medians of Fe/O related to solar wind periods from different solar sources were almost identical. These findings are in agreement with \inlinecite{lep13} who also reported relatively little variation of the Fe/O ratio over the solar cycle. Our present study further demonstrates that the Fe/O ratio does not vary significantly with the solar wind source, at least on average. 

One of the clearest temporal trends  were observed for the solar wind charge state ratio C$^{+6}/$C$^{+4}$. This ratio peaked when the active regions were present and the C$^{+6}/$C$^{+4}$ median was also clearly higher for the ARS wind than for the QS and CHS winds. These findings are consistent with \cite{fu15} who studied annual variations in the solar wind speed and the O$^{+7}/$O$^{+6}$ ratio of the slow solar wind of different origin (QS, ARS and CHS) using a ballistic two-step mapping procedure. As explained in Section 2, we also checked the results using the O$^{+7}/$O$^{+6}$ ratio and found them to be very similar  to those obtained using the C$^{+6}/$C$^{+4}$ ratio. However, \cite{fu15} found very little difference in the annual averages of O$^{+7}/$O$^{+6}$ for the QS and CHS slow solar wind during solar minimum years 2007--2008 (see their Figure 7), while we found that the QS wind had clearly the lowest C$^{+6}/$C$^{+4}$ values. Overall, our results are also consistent with \inlinecite{lep13} who found a decline in the C$^{+6}$/C$^{+4}$ ratio with decreasing solar activity. Our study suggests that this decline could be related to the decrease in the active region contribution to the slow solar wind. 

Both our study and \cite{fu15} reported that the fastest slow solar wind comes from the CHS regions. This is expected as the CHS slow wind includes the transition from the slow wind towards the fast wind. In contrast, \cite{fu15} found very similar speeds for the QS and ARS wind, while our study shows that the QS wind was clearly the slowest. The slowest speeds during our study period occurred during the deepest minimum in 2009 when the neutral line of the HCS was flat and had low tilt suggesting a large contribution to the ecliptic slow solar wind from the streamer belt region. This unusually slow solar wind period has been reported in several other works as well \citep[e.g.,][]{jia11,tsu11,kil14}. We also found that the rate of small solar wind transients peaked at this time. The high density tail (and the tendency for low entropies) for the CHS wind may be due to the contribution of the heliospheric plasma sheet (HPS) \citep[e.g.,][]{bav97}. The HPS corresponds to the stalks of the helmet streamers \citep[e.g.,][]{woo97} and this high-density structure regularly precedes slow-fast stream transition regions \citep[e.g.,][]{jia09}. 

Some previous studies suggests a negative correlation between the solar wind speed and the He/H ratio, which arises from the changes in the magnetic field expansion factor  and the efficiency of the Coulomb drag \citep[e.g.,][]{ael01,glo03,kas07}. Such correlation would  imply that the CHS related slow wind had the largest He/H values. In our study the Carrington rotation medians of the He/H ratio indeed peaked during the first half of 2008 when the fastest solar wind occurred. However, the distributions showed a clear tendency for the ARS related wind to have highest  He/H values.  As discussed in \inlinecite{lep13} the correlation between the solar wind speed and He/H values is not at all clear. Low He/H values are in particular reported near solar wind sector boundaries \citep{bor81}. Hence, low He/H values for the CHS and QS wind in our study could be related to the crossings of the HCS.

Consistent with \cite{fu15}, our study suggests that the quiet Sun has a major contribution to the slow solar wind throughout the solar minimum. This is an expected result as at this time active regions as well as low-latitude coronal holes and/or near-equatorial coronal hole extensions were absent. The origin of the slow solar wind in the QS has been studied by \cite{fel05} using elemental abundances and freeze-in temperatures as tracers to locate the source regions. The authors speculated that closed loops with lifetime of 1--2 days can possibly be the contributor to the slow solar wind streams. Their observational analysis identified only larger loops (up to 150\,000~$km$ long) as a source region. \citet{he2010} put forward the idea that the slow solar wind is initiated at the corona formation temperature of Fe~{\sc xii} at 1.6$\times$10$^6$~K. The slow solar wind plasma can be produced from magnetic reconnection between the magnetic loops and open field lines of funnels rooted in the network. The plasma released from the closed loops during the reconnection process will supply to the initial outflow of the slow solar wind. As concluded by \citet{he2010} ``there is still a long way to a full understanding of the ultimate origin of solar wind in the CH and in the QS". 

Our results highlight the difficulty in distinguishing between the slow solar wind of different origin based on the inspection of the solar wind conditions. However, some clear tendencies could be drawn. 
The QS slow solar wind stands out as slowest and coldest with the lowest values of specific entropy, C$^{+6}$/C$^{+4}$ and He/H. In contrast, the solar wind associated with the ARS was the most tenuous with the largest He/H and C$^{+6}$/C$^{+4}$ ratio. The fastest slow solar wind with the highest temperatures and entropies was related to CHS. According to our study the FIP bias (Fe/O) is not different, at least in the average sense, for different solar wind sources, and hence, cannot be used a reliable discriminator for the solar wind source. However, we would like to emphasize that our results are statistical and the slow solar wind often exhibits large variations over relatively short periods of time. To establish a more detailed linkage would require a more dedicated connection of the solar wind periods to their solar sources using advanced methods such as the non-linear force free field (NLFFF) reconstructions.

\acknowledgements
Academy of Finland project  1218152 is thanked for financial support. MM and EK acknowledge with gratitude the Royal Society  international exchange grant for the project  ``Coupling transient activity from the Sun to the Heliosphere''. MM is supported by  the Leverhulme Trust. The OMNI data were obtained through the NSSDC CDAWEB online facility. We acknowledge ACE Science Center for providing SWICS measurements. SOHO is a project of international cooperation between ESA and NASA.
\clearpage


\begin{table*}
\begin{center}  
\small  
\begin{tabular}{l l l l l}   
Parameter  & QS (LQR/\textbf{Median}/UQR) & ARS   & CHS   & ST  \\   
 \hline   
$V_{SW}$ [km/s]      & 29/\textbf{340}/36                 & 31/\textbf{363}/41                 & 49/\textbf{397}/24                    & 29/\textbf{370}/50      \\     
$n_p$ [cm$^{-3}$]    &  1.70/\textbf{5.90}/2.90         &  1.30/\textbf{4.90}/2.90        & 1.90/\textbf{6.30}/4.90              & 2.02/\textbf{7.23}/3.87            \\
$T_p$ [eV]               &  1.22/\textbf{3.05}/2.21         &  1.36/\textbf{3.61}/2.36          & 1.58/\textbf{4.24}/1.95            & 2.34/\textbf{6.41}/3.74   \\
He/H                        &  0.0060/\textbf{0.014}/0.010  &  0.0070/\textbf{0.021}/0.011  & 0.0050/\textbf{0.014}/0.0080  & 0.0092/\textbf{0.018}/0.012   \\
Fe/O                         &  0.036/\textbf{0.14}/0.055     &  0.035/\textbf{0.16}/0.042     & 0.038/\textbf{0.14}/0.050         & 0.011/\textbf{0.13}/0.050   \\
C$^{+6}/$C$^{+4}$  &  0.74/\textbf{1.89}/1.19          &  1.06/\textbf{2.77}/2.11         & 1.24/\textbf{2.29}/1.72            & 1.98/\textbf{1.86}/1.18    \\
Entropy                     &  0.61/\textbf{2.69}/0.63         &  0.53/\textbf{2.95}/0.62        & 0.86/\textbf{3.12}/0.40             & 0.66/\textbf{3.28}/0.52   \\
\hline
\end{tabular}   
\caption{Medians (bold) of selected solar wind parameters for the slow solar wind ($< 450 km s^{-1}$) related to the Quiet Sun (QS), Active Region Sun (ARS),  the Coronal Hole Sun (CHS), and the small transients (STs). For the definition of QS, ARS, and CHS periods see Section~2. The ST statistics is  shown in Figure~6.  LQR is the lower quartile range and UQR -- the upper quartile range. }
\label{tab:stealth}
\end{center}
\end{table*}





\clearpage

\bibliographystyle{spr-mp-sola}
\bibliography{ref.bib}



%
%

\clearpage

\end{article}
\end{document}